%% file: main.tex
\documentclass[conference]{IEEEtran}

\IEEEoverridecommandlockouts

\usepackage{amsmath,amsthm,relsize}
\usepackage{amsfonts, amssymb, cuted}
\usepackage{cleveref}
\usepackage{mathtools}
\usepackage{algorithm}
\usepackage[noend]{algpseudocode}
\usepackage{cite}
\usepackage{xcolor}
\usepackage{soul}
\usepackage{graphicx}
\usepackage{tikz}
\usepackage{pgfplotstable}
\usepackage{pgfplots}
\usepackage{nicefrac}
\usepackage{enumitem}
\usepackage{support-caption}
\usepackage{subcaption}
\usepackage[font=footnotesize]{caption}
\usepackage{multirow}
\pgfplotsset{compat=1.15}
\usepackage{relsize}
\usetikzlibrary{patterns}
\usepackage{acro}
\usepackage{todonotes}

\input{commands}

\input{abbrev}
\input{MyCommands}

\allowdisplaybreaks

\title{Optimal Fairness Scheduling for Coded Caching\\in Multi-AP Multi-antenna WLAN}

\begin{document}

\author{\IEEEauthorblockN{Kagan Akcay\IEEEauthorrefmark{1}, MohammadJavad Salehi\IEEEauthorrefmark{2}, Antti T\"olli\IEEEauthorrefmark{2}, and Giuseppe Caire\IEEEauthorrefmark{1}} \vspace{-5pt} \\
\IEEEauthorblockA{
    \IEEEauthorrefmark{1} Electrical Engineering and Computer Science Department, Technische Universit\"at Berlin, 10587 Berlin, Germany\\
    \IEEEauthorrefmark{2} Centre for Wireless Communications (CWC), University of Oulu, 90570 Oulu, Finland \\
    \textrm{kagan.akcay@tu-berlin.de \quad mohammadjavad.salehi@oulu.fi \quad antti.tolli@oulu.fi \quad caire@tu-berlin.de}
    \vspace{-17pt}
    }
\thanks{
The work of G. Caire and K. Akcay was partially funded by the European Research Council under the ERC Advanced Grant N. 789190, CARENET.}
}

\maketitle

\begin{abstract}
Coded caching (CC) schemes exploit the cumulative cache memory of network users, outperforming traditional uncoded schemes where cache contents are only used locally. Interestingly, this CC gain can also be combined with the spatial multiplexing gain of multi-antenna transmissions. In this paper, we extend the existing results of CC-aided data delivery in multi-access point (AP) wireless local area networks (WLAN) and video streaming applications by assuming multi-antenna transmitters at AP nodes. We present two distinct methods for using the extra resource that multi-antenna transmitters provide.
While the first method tries to reduce the number of interference links in the network graph, the second one aims to remove inter-stream interference so that users with similar cache contents can be served simultaneously. While both methods provide increased throughput, they differ significantly in the underlying concept. Numerical simulations are used to compare the performance of different methods.
\end{abstract}


%\section*{Extended Summary}
%\input{Extended_Summary}

\vspace{-10pt}
\section{Introduction}
\label{section:intro}

\input{Intro}

\section{System Model}
\label{section:sys_model}
\input{SysModel}

\section{Data Transmission in SISO Networks}
%%\subsection{Analytical Solutions}
%\label{section:optimal_solution}
\label{section:siso}
\input{SISO.tex}

\section{Data Transmission in MISO Networks}
\label{section:miso}
\input{MISO}

\section{Numerical Results}
\label{section:sim_results}
\input{SimResults.tex}

\section{Conclusion and Future Work}
We extended the existing results of cache-aided data delivery in multi-access point (AP) wireless local area networks (WLAN) and video streaming applications by assuming multi-antenna transmitters at AP nodes. We proposed two distinct methods, interference reduction and cache congestion control, to exploit the spatial multiplexing gain of multiple antennas. We used numerical simulations to analyze the performance of the two methods and compared them with the single antenna case. It was shown that cache congestion control generally performed better than interference reduction, and both methods outperformed the single antenna case by a wide margin.

It was assumed that all APs exploit the same transmission strategy, and the spatial multiplexing gain was limited to two.  
%We assumed throughout the paper that every helper uses the same method at the same scheduling decision, i.e., either Interference Nulling or Cache Congestion Control. 
In reality, each AP can select its transmission strategy independently, 
%any of the methods at the same scheduling decision. For simplicity, we have also set the spatial multiplexing gain to $2$. For higher 
and if the spatial multiplexing gain is large, a mixture of different strategies at the same AP is also possible in a single time slot. %every helper can also use a combination of the two methods at every scheduling decision. 
These options significantly expand the throughput region and will be investigated in our future works.

\bibliographystyle{IEEEtran}
\bibliography{references,kagan_references}

%\clearpage
%\newpage
%\appendix
%\input{Appendix.tex}

\end{document}

%% file: commands.tex
\newcommand{\vbar}{\raisebox{.17ex}{\rule{.04em}{1.35ex}}}
\newcommand{\vbarind}{\raisebox{.01ex}{\rule{.04em}{1.1ex}}}
\newcommand{\R}{\ifmmode{\rm I}\hspace{-.2em}{\rm R} \else ${\rm I}\hspace{-.2em}{\rm R}$ \fi}
\newcommand{\T}{\ifmmode{\rm I}\hspace{-.2em}{\rm T} \else ${\rm I}\hspace{-.2em}{\rm T}$ \fi}
\newcommand{\N}{\ifmmode{\rm I}\hspace{-.2em}{\rm N} \else \mbox{${\rm I}\hspace{-.2em}{\rm N}$} \fi}
\newcommand{\B}{\ifmmode{\rm I}\hspace{-.2em}{\rm B} \else \mbox{${\rm I}\hspace{-.2em}{\rm B}$} \fi}
\newcommand{\Hil}{\ifmmode{\rm I}\hspace{-.2em}{\rm H} \else \mbox{${\rm I}\hspace{-.2em}{\rm H}$} \fi}
\newcommand{\C}{\ifmmode\hspace{.2em}\vbar\hspace{-.31em}{\rm C} \else \mbox{$\hspace{.2em}\vbar\hspace{-.31em}{\rm C}$} \fi}
\newcommand{\Cind}{\ifmmode\hspace{.2em}\vbarind\hspace{-.25em}{\rm C} \else \mbox{$\hspace{.2em}\vbarind\hspace{-.25em}{\rm C}$} \fi}
\newcommand{\Q}{\ifmmode\hspace{.2em}\vbar\hspace{-.31em}{\rm Q} \else \mbox{$\hspace{.2em}\vbar\hspace{-.31em}{\rm Q}$} \fi}
\newcommand{\Z}{\ifmmode{\rm Z}\hspace{-.28em}{\rm Z} \else ${\rm Z}\hspace{-.28em}{\rm Z}$ \fi}

%% file: abbrev.tex
\DeclareAcronym{AWGN}{
    short = AWGN,
    long = additive white Gaussian noise,
    list = Additive White Gaussian Noise,
    tag = abbrev
}

\DeclareAcronym{ADMM}{
    short = ADMM,
    long = alternating direction method of multipliers,
    list = Alternating Direction Method of Multipliers,
    tag = abbrev
}

\DeclareAcronym{MGMC}{
    short = MGMC,
    long = multi-group multi-casting,
    list = multi-group multi-casting,
    tag = abbrev
}

\DeclareAcronym{SGMC}{
    short = SGMC,
    long = single-group multi-casting,
    list = single-group multi-casting,
    tag = abbrev
}
\DeclareAcronym{AoA}{
    short = AoA,
    long = angle-of-arrival,
    list = Angle-of-Arrival,
    tag = abbrev
}

\DeclareAcronym{AoD}{
    short = AoD,
    long = angle-of-departure,
    list = Angle-of-Departure,
    tag = abbrev
}

\DeclareAcronym{KKT}{
    short = KKT,
    long = Karush-Kuhn-Tucker,
    list = Karush-Kuhn-Tucker,
    tag = abbrev
}

\DeclareAcronym{MMF}{
    short = MMF,
    long = max-min-fairness,
    list = max-min-fairness,
    tag = abbrev
}

\DeclareAcronym{WMMF}{
    short = WMMF,
    long = weighted max-min-fairness,
    list = max-min-fairness,
    tag = abbrev
}

\DeclareAcronym{BB}{
    short = BB,
    long = base band,
    list = Base Band,
    tag = abbrev
}

\DeclareAcronym{BC}{
    short = BC,
    long = broadcast channel,
    list = Broadcast Channel,
    tag = abbrev
}

\DeclareAcronym{BS}{
    short = BS,
    long = base station,
    list = Base Station,
    tag = abbrev
}

\DeclareAcronym{BR}{
    short = BR,
    long = best response,
    list = Best Response, 
    tag = abbrev
}

\DeclareAcronym{CB}{
    short = CB,
    long = coordinated beamforming,
    list = Coordinated Beamforming,
    tag = abbrev
}

\DeclareAcronym{CC}{
    short = CC,
    long = coded caching,
    list = Coded Caching,
    tag = abbrev
}

\DeclareAcronym{CE}{
    short = CE,
    long = channel estimation,
    list = Channel Estimation,
    tag = abbrev
}

\DeclareAcronym{CoMP}{
    short = CoMP,
    long = coordinated multi-point transmission,
    list = Coordinated Multi-Point Transmission,
    tag = abbrev
}

\DeclareAcronym{CRAN}{
    short = C-RAN,
    long = cloud radio access network,
    list = Cloud Radio Access Network,
    tag = abbrev
}

\DeclareAcronym{CSE}{
    short = CSE,
    long = channel specific estimation,
    list = Channel Specific Estimation,
    tag = abbrev
}

\DeclareAcronym{CSI}{
    short = CSI,
    long = channel state information,
    list = Channel State Information,
    tag = abbrev
}

\DeclareAcronym{CSIT}{
    short = CSIT,
    long = channel state information at the transmitter,
    list = Channel State Information at the Transmitter,
    tag = abbrev
}

\DeclareAcronym{CU}{
    short = CU,
    long = central unit,
    list = Central Unit,
    tag = abbrev
}

\DeclareAcronym{D2D}{
    short = D2D,
    long = device-to-device,
    list = Device-to-Device,
    tag = abbrev
}

\DeclareAcronym{DE-ADMM}{
    short = DE-ADMM,
    long = direct estimation with alternating direction method of multipliers,
    list = Direct Estimation with Alternating Direction Method of Multipliers,
    tag = abbrev
}

\DeclareAcronym{DE-BR}{
    short = DE-BR,
    long = direct estimation with best response,
    list = Direct Estimation with Best Response,
    tag = abbrev
}

\DeclareAcronym{DE-SG}{
    short = DE-SG,
    long = direct estimation with stochastic gradient,
    list = Direct Estimation with Stochastic Gradient,
    tag = abbrev
}

\DeclareAcronym{DFT}{
	short = DFT,
	long = discrete fourier transform,
	list = Discrete Fourier Transform,
	tag = abbrev
}

\DeclareAcronym{DoF}{
    short = DoF,
    long = degrees of freedom,
    list = Degrees of Freedom,
    tag = abbrev
}

\DeclareAcronym{DL}{
    short = DL,
    long = downlink,
    list = Downlink,
    tag = abbrev
}

\DeclareAcronym{GD}{
	short = GD, 
	long = gradient descent,
	list = Gradeitn Descent,
	tag = abbrev
}

\DeclareAcronym{IBC}{
    short = IBC,
    long = interfering broadcast channel,
    list = Interfering Broadcast Channel,
    tag = abbrev
}

\DeclareAcronym{i.i.d.}{
    short = i.i.d.,
    long = independent and identically distributed,
    list = Independent and Identically Distributed,
    tag = abbrev
}

\DeclareAcronym{JP}{
    short = JP,
    long = joint processing,
    list = Joint Processing,
    tag = abbrev
}

\DeclareAcronym{LOS}{
	short = LOS,
	long = line-of-sight,
	list = Line-of-Sight,
	tag = abbrev
}

\DeclareAcronym{LS}{
    short = LS,
    long = least squares,
    list = Least Squares,
    tag = abbrev
}

\DeclareAcronym{LTE}{
    short = LTE,
    long = Long Term Evolution,
    tag = abbrev
}

\DeclareAcronym{LTE-A}{
    short = LTE-A,
    long = Long Term Evolution Advanced,
    tag = abbrev
}

\DeclareAcronym{MIMO}{
    short = MIMO,
    long = multiple-input multiple-output,
    list = Multiple-Input Multiple-Output,
    tag = abbrev
}

\DeclareAcronym{MISO}{
    short = MISO,
    long = multiple-input single-output,
    list = Multiple-Input Single-Output,
    tag = abbrev
}

\DeclareAcronym{MAC}{
    short = MAC,
    long = multiple access channel,
    list = Multiple Access Channel,
    tag = abbrev
}

\DeclareAcronym{MSE}{
    short = MSE,
    long = mean-squared error,
    list = Mean-Squared Error,
    tag = abbrev
}

\DeclareAcronym{MMSE}{
    short = MMSE,
    long = minimum mean-squared error,
    list = Minimum Mean-Squared Error,
    tag = abbrev
}

\DeclareAcronym{mmWave}{
	short = mmWave,
	long = millimeter wave,
	list = Millimeter Wave,
	tag = abbrev
}

\DeclareAcronym{MU-MIMO}{
    short = MU-MIMO,
    long = multi-user \ac{MIMO},
    list = Multi-User \ac{MIMO},
    tag = abbrev
}

\DeclareAcronym{OTA}{
    short = OTA,
    long = over-the-air,
    list = Over-the-Air,
    tag = abbrev
}

\DeclareAcronym{PSD}{
    short = PSD,
    long = positive semidefinite,
    list = Positive Semidefinite,
    tag = abbrev
}

\DeclareAcronym{QoS}{
	short = QoS,
	long = quality of service,
	list = Quality of Service,
	tag = abbrev
}

\DeclareAcronym{RCP}{
	short = RCP,
	long = remote central processor,
	list = Remote Central Processor,
	tag = abbrev
}

\DeclareAcronym{RRH}{
    short = RRH,
    long = remote radio head,
    list = Remote Radio Head,
    tag = abbrev
}

\DeclareAcronym{RSSI}{
    short = RSSI,
    long = received signal strength indicator,
    list = Received Signal Strength Indicator,
    tag = abbrev
}

\DeclareAcronym{RX}{
	short = RX,
	long = receiver,
	list = Receiver,
	tag = abbrev
}

\DeclareAcronym{SCA}{
    short = SCA,
    long = successive-convex-approximation,
    list = Successive-Convex-Approximation,
    tag = abbrev
}

\DeclareAcronym{SG}{
    short = SG,
    long = stochastic gradient,
    list = Stochastic Gradient,
    tag = abbrev
}

\DeclareAcronym{SIC}{
    short = SIC,
    long = successive interference cancellation,
    list = Successive Interference Cancellation,
    tag = abbrev
}

\DeclareAcronym{SNR}{
    short = SNR,
    long = signal-to-noise-ratio,
    list = Signal-to-Noise Ratio,
    tag = abbrev
}

\DeclareAcronym{SDR}{
    short = SDR,
    long = semi-definite-relaxation,
    list = semi-definite-relaxation,
    tag = abbrev
}

\DeclareAcronym{SINR}{
    short = SINR,
    long = signal-to-interference-plus-noise ratio,
    list = Signal-to-Interference-plus-Noise Ratio,
    tag = abbrev
}

\DeclareAcronym{SOCP}{
	short = SOCP, 
	long = second order cone program,
	list = Second Order Cone Program,
	tag = abbrev
}

\DeclareAcronym{SSE}{
    short = SSE,
    long = stream specific estimation,
    list = Stream Specific Estimation,
    tag = abbrev
}

\DeclareAcronym{SVD}{
	short = SVD,
	long = singular value decomposition,
	list = Singular Value Decomposition,
	tag = abbrev
}

\DeclareAcronym{TDD}{
	short = TDD,
	long = time division duplex,
	list = Time Division Duplex,
	tag = abbrev
}

\DeclareAcronym{TX}{
	short = TX,
	long = transmitter,
	list = Transmitter,
	tag = abbrev
}

\DeclareAcronym{UE}{
    short = UE,
    long = user equipment,
    list = User Equipment,
    tag = abbrev
}

\DeclareAcronym{UL}{
    short = UL,
    long = uplink,
    list = Uplink,
    tag = abbrev
}

\DeclareAcronym{ULA}{
	short = ULA,
	long = uniform linear array,
	list = Uniform Linear Array,
	tag = abbrev
}

\DeclareAcronym{UPA}{
    short = UPA,
    long = uniform planar array,
    list = Uniform Planar Array,
    tag = abbrev
}

\DeclareAcronym{WMMSE}{
    short = WMMSE,
    long = weighted minimum mean-squared error,
    list = Weighted Minimum Mean-Squared Error,
    tag = abbrev
}

\DeclareAcronym{WMSEMin}{
    short = WMSEMin,
    long = weighted sum \ac{MSE} minimization,
    list = Weighted sum \ac{MSE} Minimization,
    tag = abbrev
}

\DeclareAcronym{WBAN}{
	short = WBAN,
	long = wireless body area network,
	list = Wireless Body Area Network,
	tag = abbrev
}

\DeclareAcronym{WSRMax}{
    short = WSRMax,
    long = weighted sum rate maximization,
    list = Weighted Sum Rate Maximization,
    tag = abbrev
}

%% file: MyCommands.tex
\newtheorem{exmp}{Example}%[section]
\theoremstyle{definition}

\newcommand{\CA}[0]{{\mathcal{A}}}
\newcommand{\CB}[0]{{\mathcal{B}}}

\newcommand{\CH}[0]{{\mathcal{H}}}
\newcommand{\CI}[0]{{\mathcal{I}}}

\newcommand{\CL}[0]{{\mathcal{L}}}
\newcommand{\CM}[0]{{\mathcal{M}}}

\newcommand{\CS}[0]{{\mathcal{S}}}
\newcommand{\CT}[0]{{\mathcal{T}}}
\newcommand{\CU}[0]{{\mathcal{U}}}
\newcommand{\CV}[0]{{\mathcal{V}}}

\newcommand{\CZ}[0]{{\mathcal{Z}}}

\newcommand{\Bp}[0]{{\mathbf{p}}}

\newcommand{\Br}[0]{{\mathbf{r}}}

\newcommand{\Bw}[0]{{\mathbf{w}}}

\newcommand{\Sfc}[0]{{\mathsf{c}}}

\newcommand{\Sfn}[0]{{\mathsf{n}}}

\newcommand{\SfL}[0]{{\mathsf{L}}}

%% file: Intro.tex
The increasing amount of data traffic, especially driven by multimedia applications, has necessitated more efficient usage of existing network resources and exploring new possibilities. One interesting resource is on-device memory; it is cheap and can be used to store a large part of multimedia content proactively. In this regard, coded caching (CC) techniques provide an exciting opportunity to use memory as a communication resource, as they enable a speedup factor in the achievable rate that scales with the cumulative cache size in the network~\cite{maddah2014fundamental} and is additive with the spatial multiplexing gain of multi-antenna communications~\cite{shariatpanahi2018physical,parrinellomultiantenna}.

%, e.g., for video-on-demand (VoD) and extended reality (XR) applications. Many researchers have considered the efficient use of onboard memory in the context of \emph{caching}, with pioneering works introducing femtocaching~\cite{shanmugam2013femtocaching} and F-RAN models~\cite{park2016joint}. A major breakthrough later happened with the introduction of coded caching (CC)~\cite{maddah2014fundamental}, which showed a speedup factor in the achievable rate, scaling with the cumulative cache size in the network, is possible by multicasting carefully created codewords. Later, many works extended the original work of~\cite{maddah2014fundamental}, e.g., for wireless~\cite{tolli2017multi}, multi-server~\cite{shariatpanahi2016multi}, and multi-antenna~\cite{shariatpanahi2018physical,parrinellomultiantenna} networks. Meantime, many challenges hindering the applicability of CC techniques in real networks were addressed. For example, the subpacketization bottleneck was studied for single- and multi-antenna setups~\cite{lampiris2018adding,salehi2020lowcomplexity}, simpler beamformer designs were introduced~\cite{salehi2020lowcomplexity,mahmoodi2021low}, and the problem of supporting network dynamicity was addressed~\cite{caire,salehi2021low,abolpour2022coded}.

This paper considers a coded caching system model originally proposed in~\cite{mozhgan} and later extended in~\cite{akcay2023optimal} for two-hop wireless local area networks (WLAN), where a single server transmits data to a large number of users through multiple access points (AP), hereby called \textit{helper} nodes.
%WLANs play an essential role in our everyday lives; they facilitate data access, e.g., in our homes, malls, airports, and cars. 
Two important properties of WLANs are a) unit frequency reuse due to the scarce availability of unlicensed bands and b) the collision-type interference model (i.e., packets are lost if a receiver gets the superposition of concurrent packets, all above a certain interference threshold, from different helpers). Considering multi-helper WLANs and single per-user requests, the authors in~\cite{mozhgan} 
used the multi-round delivery method in~\cite{caire} (which was later proved by~\cite{parrinellomultiantenna} to be optimal for the shared-cache model) and graph coloring to construct a reuse pattern of helpers and optimize the helper-user association to minimize the delivery time.
%optimized the worst-case delivery time of all requests. Using basic principles of the multi-round delivery method in~\cite{caire} (which was later proved by~\cite{parrinellomultiantenna} to be optimal for the shared-cache model), they first used graph coloring to construct a reuse pattern of helpers and optimized the helper-user association to minimize the delivery time. Next, they introduced the heuristic `Avalanche' algorithm to leverage the interference-free state of the users as helpers finished their multi-round delivery. 
This work was later extended in~\cite{akcay2023optimal}, where the authors developed a computational method to determine the theoretical throughput region of the users' content delivery rates (defined as the number of chunks delivered per unit of time per user) and solved the fairness scheduling problem by maximizing the desired fairness metric over this throughput region. 
%Two low-complexity heuristic methods were also introduced in~[ISIT], where one maximized the desired fairness metric over a sub-region of the entire throughput region, and the other used a greedy algorithmic approach to fairly associate users with helpers.

Both~\cite{mozhgan},~\cite{akcay2023optimal} consider single-input single-output (SISO) setups only. Given the undoubted importance of multi-antenna communications~\cite{rajatheva2020white}, this paper provides an extension of~\cite{akcay2023optimal} to multi-input single-output (MISO) setups where multi-antenna helpers communicate with single-antenna users. Specifically, we recognize two broad methods to use the additional networking resource provided by the MISO extension. % The first method tries to reduce the number of interference links in the network graph, and the second one aims to remove inter-stream interference so that users with similar cache contents can be served simultaneously. %\textbf{Until here introduction seems generally fine, we can then here explain the two methods better in  couple of sentences. We may add notation part also if we have space.}
In the first method, called \emph{interference reduction}, we assume that at each helper, appropriate beamformers are used to suppress the transmitted signal at a well-defined subset of users in the interference range of the helper, enabling these users to be served by other nearby helpers simultaneously. On the other hand, with the other method, called \emph{cache congestion control}, we assume beamformers are designed to remove the inter-stream interference at users with the same cache content (which is a natural result of the cache placement of~\cite{caire}), enabling these users to be served simultaneously in a similar way to the shared-cache model~\cite{parrinello2019fundamental,parrinello2020extending}. 
Numerical simulations are used to compare the performance of these two methods.
%While both methods enable serving more users with each transmission, they differ significantly in the underlying concept and resulting performance. In this extended summary, we provide a simplified definition of the system model, the intuition behind each method, and preliminary simulation results comparing them. More detailed explanations are due for the extended version. 

In the following, we have used boldface lower-case letters to denote vectors. $\Br[i]$ is the $i$-th element of vector $\Br$. Calligraphic letters represent sets, and $\CA \backslash \CB$ is the set of elements of $\CA$ not in $\CB$. $[L]$ represents the set $\{1,2,\cdots,L\}$. %Other notations are defined as used in the text.

%In the following, by $[L]$, we mean the set of numbers $\{1,\cdots,L\}$.

%

%% file: SysModel.tex
The system model is similar to the broadcast/collision model of~\cite{akcay2023optimal}: a single server, connected to $H$ multi-antenna helper nodes, serves the requests of $K$ cache-enabled single-antenna users. We use $h_i$ and $u_k$, $i \in [H]$ and $k \in [K]$, to represent helpers and users, respectively. Each helper $h_i$ is capable of attaining the spatial multiplexing gain of $\alpha$ (the actual number of antennas may be larger).
%
%
%For the network setup, we consider $H$ multi-antenna helper nodes, each capable of attaining the spatial multiplexing gain of $\alpha$ (the actual number of antennas may be larger), serving the requests of $K$ single-antenna users. 
The transmission and interference radii of all helpers are the same.
%Every helper $h_i$ has a transmission radius of $r_{\textrm{trans}}$ and an interference radius of $r_{\mathrm{inter}} \ge r_{\mathrm{trans}}$, and may be either active or inactive (i.e., not transmitting any signal) at any time interval. 
A user $u_k$ can successfully decode the message transmitted by the active helper $h_i$ if 1) it is within the transmission radius of $h_i$, and 2) it is not within the interference radius of any other active helper $h_{i'}$, unless the interference caused by $h_{i'}$ on $u_k$ is suppressed by beamforming at $h_{i'}$. 
%A graphical representation of the system model is shown in Figure~\ref{fig:network_model}.
%
Like~\cite{akcay2023optimal}, we consider a video streaming application where every video file is split into multiple `chunks,' each representing a few seconds of the video, and every user generates a sequence of requests for the chunks of the video it is streaming. All chunks have the same size, and every user's cache memory can store a portion $\gamma$ of all the chunks. System operation consists of two phases, placement and delivery. 
%During the placement phase, users' cache memories are filled up with subpackets (i.e., smaller portions) of all the video chunks. 
Similar to~\cite{akcay2023optimal}, to keep the subpacketization level low, we use the decentralized cache replication algorithm of~\cite{caire,parrinellomultiantenna} for the placement phase: we fix a number of $L < K$ \emph{cache profiles} and assign each user randomly to one of the profiles. 
%Let us use $\SfL(u_k) \in [L]$ to denote the cache profile assigned to user $u_k$. Given $L$, 
Defining the CC gain as $t = L \gamma$, we divide every video chunk $W$ into $\binom{L}{t}$ non-overlapping equal-sized subpackets $W_{\CS}$, where $\CS$ can be any subset of $[L]$ with $|\CS| = t$. Then, if a user $u_k$ is assigned to profile $l \in [L]$, we store every subpacket $W_{\CS}$, for every chunk $W$ and every $\CS \ni l$, in the cache memory of $u_k$ (clearly, all users assigned to the same profile will cache the same data).
Accordingly, in the delivery phase, after users reveal their requested video chunks, the server creates a number of transmission vectors and transmits them through a selection of active helpers. 
%We assume users can start streaming at arbitrary times, and when they start streaming, they repeatedly generate requests for consecutive video chunks. With this model, we can safely assume a single chunk is requested by at most one user at a given time, as even if two users are streaming the same video, it is unlikely that they have started streaming at the same time. Given the set of user requests and using a delivery algorithm, 
%The server then creates a number of transmission vectors and transmits them through a selection of active helpers. The goal is to serve requested video chunks while optimizing a performance function of, e.g., delivery time, energy, or fairness.
%
With $H$ helpers that can be either active or inactive, there exist $2^H-1$ selections for the \emph{activation pattern} of the helpers. Let us use vectors $\Bp_j$, $j \in [2^H-1]$, with binary elements, to denote activation patterns.

%Every $\Bp_j$ has $H$ binary elements $\Bp_j[i]$, $i \in [H]$, and $\Bp_j[i] = 1$ only if helper $h_i$ is active in the corresponding activation pattern. 
As discussed in~\cite{akcay2023optimal}, the activation pattern uniquely defines the set of users $\CU_i$ that could be potentially served by each active helper. 
%Let us use $\CU_i$ to denote the set of users that can be served by helper $h_i$. 
%According to the cache placement strategy, we may have multiple users with the same cache profile (hence, the same cache content). However, 
However, given the cache placement strategy, the codewords can be built only for users in the same $\CU_i$ that are assigned to different cache profiles, and hence, we may have multiple choices for codeword creation for every active helper. Let's call every possible selection of codewords for all active helpers in an activation pattern $\Bp_j$ a \emph{scheduling decison} and denote it by $(j,s)$. A scheduling decision $(j,s)$ corresponds to the selection of 
activation pattern $\Bp_j$ and codeword choice $s \in [\Sfc(\Bp_j)]$ where $\Sfc(\Bp_j)$ denotes the number of such choices for a given activation pattern $\Bp_j$. %Given a scheduling decision $(j,s)$, every user $k$ obtains an instantaneous rate $r_k(j,s)$. 
Each scheduling decision $(j,s)$ results in an instantaneous rate vector $\Br(j,s)$, where 
the element $k$, $k \in [K]$, in $\Br(j,s)$ is the number of video chunks per unit time obtained by user~$u_k$ under decision $(j,s)$. A scheduler chooses a sequence of scheduling decisions $(j,s)$, each for a fraction of time $a(j,s)$, over a sequence of time units. The long-term (time-averaged) throughput vector $\bar{\Br}$ %, measured in delivered video chunks per unit time, 
is given by $\bar{\Br} = \sum_{(j,s)} a(j,s) \Br(j,s)$, and the achievable throughput region $R$ is the convex hull of all instantaneous rate vectors. A canonical fairness scheduling problem consists of finding the scheduling rule (i.e., the fractions of time $a(j,s)$) that maximizes a suitable component-wise non-decreasing concave function of 
user throughputs over the region $R$~\cite{georgiadis2006resource}. In particular, the so-called proportional fairness criterion~\cite{fairness} considers the network utility function $f(\bar{\Br}) = \sum_k \log(\bar{\Br}[k])$, resulting in the following convex optimization problem:
\vspace{-3pt}
\begin{equation}
\label{eq:optimization_problem_main}
    %\begin{aligned}
        \max_{a(j,s)} f(\bar{\Br}) \quad s.t. \quad  a(j,s)\geq{0}, \: \sum_{(j,s)} a(j,s) = 1 . \\
        %& 0 \le a(j,s_j) \le a_{\max}, \quad \forall j\in[2^H], s_j \in [\Sfc(\Bp_j)],
    %\end{aligned}
\end{equation}

%Every policy $\pi$ under a given activation pattern $\Bp_j$ uniquely defines an instantaneous $K \times 1$ rate vector $\Br(\pi,\Bp_j)$, where the rate is defined as the number of codewords delivered per unit time per user. A scheduler sequentially chooses policies to continuously serve users, aiming to maximize a component-wise non-decreasing concave fairness function~\cite{fairness} over the fraction of time $a(\pi,\Bp_j)$ of the policy $(\pi,\Bp_j)$. We choose proportional fairness $f(\Br_{\mathrm{avg}})=\sum_{k \in [K]} \log(\Br_{\mathrm{avg}}[k])$ where $\Br_{\mathrm{avg}}[k]$ is the long-term average throughput of user $u_k$. So we have the following convex optimization problem:

%A scheduler uses each policy $\pi$ $a(\pi,\Bp_j)$ fraction of time by maximizing a componentwise nondecreasing fairness function.
%is to optimize the proportional fairness function (or any other member of the $\alpha$-fairness family~\cite{fairness}) over the convex hull of all rate vectors by solving
%\begin{equation}
%\label{eq:optimization_problem_main}
%    \begin{aligned}
%        \max_{a(\pi,\Bp_j)} \Sff &= \sum_{k \in [K]} \log(\Br_{\mathrm{avg}}[k]) \\
%        s.t. \qquad & \Br_{\mathrm{avg}} = \sum_{j \in [2^H]} \sum_{\pi} a(\pi,\Bp_j) \Br(\pi,\Bp_j), \\
%        & a(\pi,\Bp_j)\geq{0}, \quad \sum_{j \in [2^H]} \sum_{\pi} a(\pi,\Bp_j) = 1 . \\
%        %& 0 \le a(j,s_j) \le a_{\max}, \quad \forall j\in[2^H], s_j \in [\Sfc(\Bp_j)],
%    \end{aligned}
%\end{equation}

\begin{figure}[t]
    %\begin{subfigure}{0.3\columnwidth}
        \centering
        \includegraphics[width = 0.7\columnwidth]{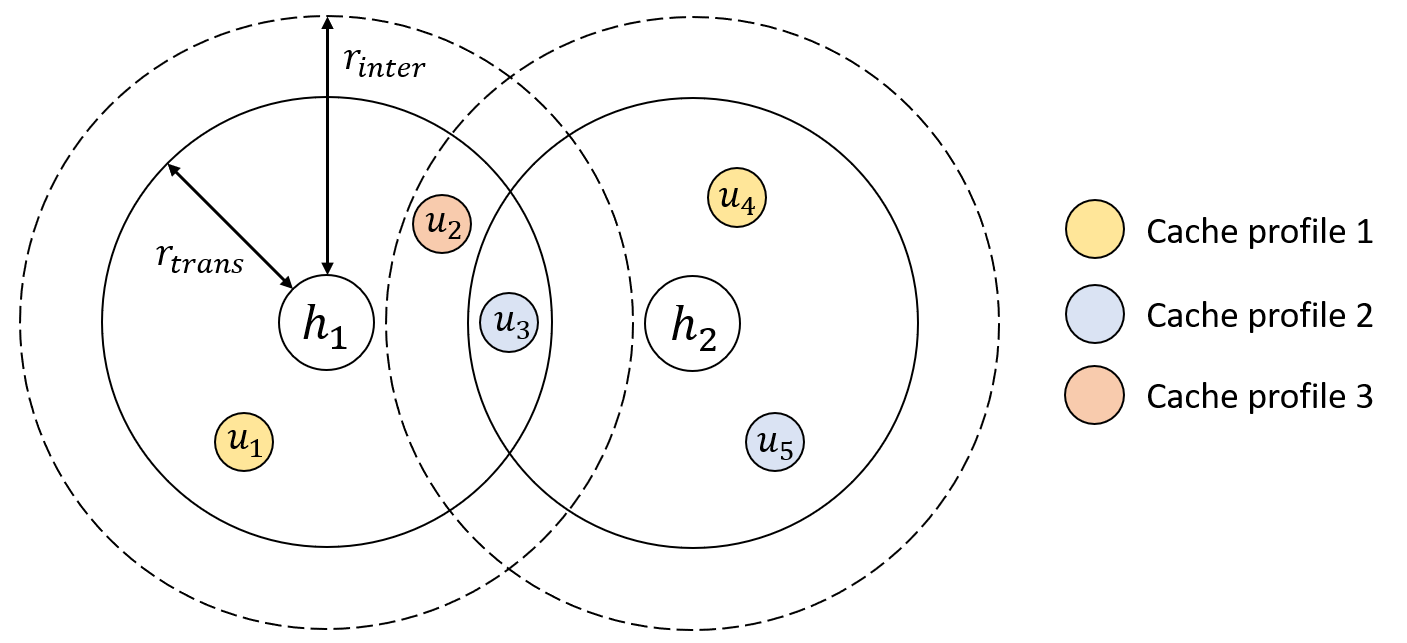} %\label{fig:network_model}
        \vspace{-1pt}
    \caption{Example network with $H=2$, $K=5$, $L=3$. Figure from~\cite{akcay2023optimal}.}
    \label{fig:network_model}
    \vspace{-12pt}
\end{figure}

%\begin{figure}[t]
    %\begin{subfigure}{0.3\columnwidth}
%        \centering
%        \includegraphics[width = 0.7\columnwidth]{Figs/NetworkModel.png} \label{fig:network_model}
%        \vspace{-1pt}
%    \caption{Example network with $H=2$, $K=5$, $L=3$. Figure from~\cite{akcay2023optimal}.}
%    \vspace{-12pt}
%    \label{fig:network_model}
%\end{figure}

%% file: SISO.tex
Assume activation pattern $\Bp_j$ is given and helper $h_i$ is active in $\Bp_j$. Let us use $r_{\mathrm{trans}}$ and $r_{\mathrm{inter}}$ to denote the transmission and interference range of active helpers. Then, the set of users that can be served by $h_i$, denoted by $\CU_i$, includes users within distance $r_{\mathrm{trans}}$ of $h_i$ but out of distance $r_{\mathrm{inter}}$ of every other active helper. We use $\CU_i^l$ to represent the subset of users in $\CU_i$ that are all assigned to the cache profile $l \in L$. Representing the profile assigned to user $u_k$ by $\SfL(u_k)$, this means that for every $u_k \in \CU_i^l$, we have $\SfL(u_k) = l$. We also use $l(i)$ to denote the number of $\CU_i^l$ sets for which $|\CU_i^l| = 0$.

The first step in codeword creation is to build a \emph{feasible} subset $\CV_i$ of $\CU_i$, where every user in $\CV_i$ is assigned to a different profile.
As in~\cite{akcay2023optimal}, throughout this paper, we restrict ourselves to feasible sets which maximize the coded caching gain, i.e., the sets $\CV_i$ with the maximum possible length so that we have $|\CV_i|=L-l(i)$.
We use $\CV_i^l$ to denote the set of users in $\CV_i$ assigned to profile $l \in [L]$ and
$\CL(\CV_i)$ to represent the set of cache profiles assigned to all users in $\CV_i$. Clearly, for the SISO case, $\CV_i^l$ is either empty or includes one user.

%, i.e., $\CL(\CV_i) = \{l \; | \; |\CU_i^l|>0, l\in{[L]}\}$. 
%$\CL(\CV_i) = \{\SfL(u_k), \forall u_k \in \CV_i \}$. 
%As in~\cite{akcay2023optimal}, throughout this paper, we restrict ourselves to feasible sets which maximize the coded caching gain, i.e., the sets $\CV_i$ with the maximum possible length so that we have $|\CV_i|=L-l(i)$. \color{black}

The users in a feasible set $\CV_i$ can be served by a number of codewords transmitted through helper $h_i$. To find these codewords, we first create the \emph{extended} set of $\CV_i$ as $\hat{\CV}_i$ by adding $L - |\CV_i|=l(i)$ \emph{phantom} users $u_{l}^*$,\footnote{Phantom users are imaginary users added only to help with the formal definition of the delivery process. The same concept is used in~\cite{salehi2020lowcomplexity}.} 
$\SfL(u_l^*) = l$, for every $l \in [L] \backslash \CL(\CV_i)$. Next, a \emph{preliminary} codeword $\hat{X}(\hat{\CT}_i)$ can be built for every subset $\hat{\CT}_i$ of $\hat{\CV}_i$ with size $|\hat{\CT}_i| = t+1$ as
\begin{equation}
\label{eq:xor_preliminary}
    \hat{X}(\hat{\CT}_i) = \bigoplus_{u_{k} \in {\hat{\CT}_i}} W_{d_{k}, \CL(\hat{\CT}_i) \backslash \{\SfL(u_k)\} } ,
\end{equation}
where $\CL(\hat{\CT}_i)$ is the set of cache profiles assigned to all users in $\hat{\CT}_i$
%\color{blue} $\CL(\hat{\CT}_i) = \{l |$ s.t. $|\hat{\CT}_i^l|>0, l\in{[L]}\}$ \color{black} 
and $d_k$ is the index of the video chunk requested by $u_k$. Finally, the codeword $X(\CT_i)$ is built from $\hat{X}(\hat{\CT}_i)$ by removing the effect of phantom users ($\CT_i$ is the set resulting by removing phantom users from $\hat{\CT}_i$).
%The codeword creation process clarifies exactly which codewords are generated for any given feasible set $\CV_i$.

A total number of $\binom{l(i)}{t+1}$  preliminary codewords $\hat{X}(\hat{\CT}_i)$ include data only for phantom users and should be completely ignored in the transmission. So, every user $u_k$ in the feasible set $\CV_i$ can successfully decode its requested chunk after $ \Sfn(\CV_i) = {\binom{L}{t+1} - \binom{l(i)}{t+1}}$ codeword transmissions, and hence, its rate can be simply calculated as $r(\CV_i) = {1}/{\Sfn(\CV_i)}$. 
%\begin{equation}
    %$r(\CV_i) = {1}/({\binom{L}{t+1} - \binom{L-|\CV_i|}{t+1}})$.
%\end{equation}
%which results from the fact that only phantom users appear in $\binom{L-|\CV_i|}{t+1}$ number of preliminary codewords $\hat{X}(\hat{\CT}_i)$. 
As a result, to find the instantaneous rate vector $\Br(j,s)$, we need to 1) for every active helper $h_i$, choose a feasible set $\CV_i$, 2) for every $u_k \in \CV_i$, set the element $k$ of the vector $\Br(j,s)$ as $r(\CV_i)$, and 3) fill other elements of $\Br(j,s)$ as zero. 
%Note that $r(\CV_i)$ is calculated according to the fact that $\binom{l(i)}{t+1}$  preliminary codewords $\hat{X}(\hat{\CT}_i)$ are ignored as they include only phantom users.

\begin{exmp}
\label{exmp:siso}
Consider the simple network in Figure~\ref{fig:network_model}. Assume $t=1$, and consider the activation pattern $\Bp_j = [0,1]$ (i.e., only helper $h_2$ is active). Given the cache configuration in the figure, we have two possible policies, which are to serve either $\{u_4,u_3\}$ or $\{u_4,u_5\}$ (note that $u_2$ is \emph{not} in the transmission radius of $h_2$). Let us consider the scheduling decision $(j,s)$ to serve $\{u_4,u_3\}$, and denote the chunks requested by $u_3$ and $u_4$ as $C$ and $D$, respectively. As $u_3$ and $u_4$ belong to cache groups $2$ and $1$, they have $\{C_2,D_2\}$ and $\{C_1,D_1\}$ in their caches, respectively. As a result, we need to send $C_1 \oplus D_2$, $C_3$, and $D_3$, in 3 time slots, through $h_2$ for $u_3$ and $u_4$ to decode $\{C_1,C_3\}$ and $\{D_2,D_3\}$, respectively. As a result, the corresponding rate vector for this scheduling decision is $\Br(j,s) = [0,0,\frac{1}{3},\frac{1}{3},0]$. %\textbf{Can we leave here like this?, i.e. check for more details?} 
%(check~\cite{akcay2023optimal} for more details). %\textbf{Until here generally system model and SISO seems fine.}
\end{exmp}

%% file: MISO.tex
In this paper, we make the simplifying assumption that `beamforming for interference suppression at any user(s) does not alter the connectivity/interference status of other users.' We consider two transmission strategies for MISO networks:

%\noindent\textbf{B.1. Interference nulling: }
\subsection{Interference Reduction}
With this strategy, after selecting the activation pattern $\Bp_j$, we first select a subset of users $\CZ_i$, with $|\CZ_i| < \alpha$, from the set of users that are in the interference radius of every active helper $h_i$, and design transmit beamformers such that the transmitted signal from $h_i$ is suppressed at every user in $\CZ_i$ (recall that the spatial multiplexing gain of $\alpha$ is attainable).
%for every active helper $h_i$ we first select ,  
Then, we repeat the same process as the SISO case in Section~\ref{section:siso}, assuming that users in every $\CZ_i$ do not receive any interference from the helper $h_i$. This improves performance by allowing more users to be served simultaneously.

\begin{exmp}
\label{ex:Inter_null_example}
Consider the network in Example~\ref{exmp:siso}, and assume $t=1$, $\alpha=2$, and $\Bp_j=[1,1]$ (i.e., both helpers are active). Let us consider $\CZ_1 = \{u_3\}$ and $\CZ_2 = \{u_2\}$. With this selection, the sets of users that can successfully get data from each helper are given as $\CU_1=\{ u_1,u_2 \}$ and $\CU_2=\{ u_3,u_4,u_5 \}$. Let us select the corresponding feasible sets that maximize the coded caching gain for each helper as $\CV_1=\{ u_1,u_2 \}$ and $\CV_2 = \{u_3, u_4\}$, which means the extended sets are given as $\hat{\CV}_1=\{ u_1,u_2,u_{2}^* \}$ and $\hat{\CV}_2=\{ u_3,u_4,u_{3}^*\}$. Then, using~\eqref{eq:xor_preliminary}, the preliminary codewords transmitted from $h_1$ are built as
\begin{equation*}
    \begin{aligned}
        \hat{X}_1(\{u_1,u_2\}) &=\left(W_{d_1,\{3\}} \oplus W_{d_2,\{1\}}\right)\Bw_{\{3\}}, \\
        \hat{X}_1(\{u_1,u_2^*\}) &=\left(W_{d_1,\{2\}} \oplus W_{d_{2^*},\{1\}}\right)\Bw_{\{3\}}, \\
        \hat{X}_1(\{u_2,u_2^*\}) &=\left(W_{d_2,\{2\}} \oplus W_{d_{2^*},\{3\}}\right)\Bw_{\{3\}},
    \end{aligned}
\end{equation*}
where the beamforming vector $\Bw_{\{3\}}$ is added to suppress the transmitted data at user $u_3$ (recall that $\CZ_1 = \{u_3\}$). Denoting the files requested by users $u_1$-$u_5$ as $A$-$E$, these preliminary codewords correspond to the transmission of $(A_3\oplus B_1) \Bw_{\{3\}}$, $A_2 \Bw_{\{3\}}$, and $B_2 \Bw_{\{3\}}$ from $h_1$. Following the same procedure, $h_2$ transmits $(C_1 \oplus D_2) \Bw_{\{2\}}$, $C_3 \Bw_{\{2\}}$, and $D_3 \Bw_{\{2\}}$, and the rate vector for the considered scheduling decision is $\Br(j,s) = [\frac{1}{3},\frac{1}{3},\frac{1}{3},\frac{1}{3},0]$.
\end{exmp}

One should note that in order to improve the performance with the interference reduction strategy, for every active helper $h_i$, we should include only those users in $\CZ_i$ that can receive data from at least another active helper (other than $h_i$). This is because the main point of interference reduction is to relieve the interference on users that are close to multiple helpers so that such users can still receive data while nearby helpers are active. This intuition was the motivation behind the selection of $\CZ_1$ and $\CZ_2$ in Example~\ref{ex:Inter_null_example}.

%Let $\CH$ be the set of active helpers in $\Bp_j$. We define the degree of user $u_k$, denoted by $\Sfd(u_k)$, as the number of helpers in the interference radius of $u_k$. For every helper $h_i\in{\CH}$, it is necessary for every user $u_k\in{\CZ_i}$ to have a degree $d(u_k)>1$ \color{blue} and there is at least one active helper $h_{i'}$, other than $h_i$, within transmission radius of $u_k$ \color{black} so that $u_k$ can receive a message from $h_{i'}\in{\CH}$. 

Let us now calculate the complexity order of the interference reduction strategy. Let $\CI_i$ denote the users $u_k$ that are: 1) within the interference radius of helper $h_i$, and 2) within the transmission radius of at least another active helper.
%where $d(u_k)>1$, \color{blue} where there is at least one active helper $h_{i'}$, other than $h_i$, within transmission radius of $u_k$ \color{black}, $\forall u_k\in{I_i}$. 
Since every active helper can suppress its transmitted message at a maximum of $\alpha-1$  users, there is a maximum of $\sum_{m=0}^{\alpha-1}\binom{|\CI_i|}{m}$ possible choices to select $\CZ_i$ for every active helper $h_i$. As a result, using $\CH_j$ to denote the set of all active helpers for a given activation pattern $\Bp_j$, we have a total number of $\prod_{h_i\in{\CH_j}}(\sum_{m=0}^{\alpha-1}\binom{|\CI_i|}{m})-1$ interference reduction strategies for $\Bp_j$, where the subtraction of one is to exclude the case where all $\CZ_i$ sets are empty. Of course, we should emphasize that this is an upper bound to the number of \emph{effective} strategies that improve the performance, as not every possible strategy will be effective (for example, consider a case where a user $u_k$ is in the interference radii of two nearby active helpers, but only one of these helpers suppresses the interference on $u_k$). The exact number of effective strategies depends on the network topology.

\begin{exmp}
\label{ex:Inter_null_example_counter}
    Consider the network in Example~\ref{exmp:siso}, and assume $t=1$, $\alpha=2$, and $\Bp_j=[1,1]$.
    According to the definition, we have $\CI_1 = \{u_3\}$ and $\CI_2 = \{u_2,u_3\}$. So, the total number of possible nulling strategies is given by
    \[
    \left(\binom{1}{0}+\binom{1}{1}\right)\left(\binom{2}{0}+\binom{2}{1}\right)-1=5.
    \]
    It can be verified that among these strategies, all are effective except the case $\CZ_1 = \CZ_2 = \{u_3\}$.
    \end{exmp}

%\color{black}

%\begin{exmp}
%Consider the same network as ...
%\end{exmp}
%A helper may need to serve several users with the same cache profile since $L<K$ and users are randomly assigned to one of $L$ cache profiles. However, users with the same cache profile can not be served with the same codewords when they request distinct files. We call this \textit{cache congestion}. %Our second method helps us to reduce cache congestion by using MISO spatial gain.

\subsection{Cache Congestion Control}
\label{cache congestion}
Beyond interference, another throughput limiting factor is the limited number of cache profiles. Since cache profiles are assigned to users randomly and users can move arbitrarily in the network, it may happen that one active helper must serve several users assigned with the same profile. In this case, these users must be served sequentially, and no CC gain is achieved (recall that only users with different cache types can be served by multicasting CC codewords). We refer to this type of impairment as ``cache congestion’’. Our second approach consists of using the MISO spatial multiplexing capability to reduce cache congestion. We define the set of users $\CU_i$ that can be served by active helper $h_i$ similar to the SISO case in Section~\ref{section:siso}, but design beamformers at $h_i$ to suppress the inter-stream interference, similar to the shared-cache and dynamic CC models~\cite{parrinellomultiantenna,parrinello2020extending,abolpour2023cache}, to allow users with similar cache contents (i.e., users assigned with the same cache profile) to be served simultaneously.
 %as follows:

In order to describe content delivery using the cache congestion control mechanism, we need a few modifications to the process used in Section~\ref{section:siso} for SISO setups. Assume activation pattern $\Bp_j$ is given, and $h_i$ is an active helper in $\Bp_j$. Similar to the SISO case, we use $\CU_i$ to denote the set of users that can receive data from $h_i$. However, the feasible set $\CV_i^c$ is now defined as a subset of $\CU_i$ where at most $\alpha$ users in $\CV_i^c$ are assigned to every cache profile $l \in [L]$. Let us again limit ourselves to the feasible sets maximizing the coded caching gain. Then, we have $|\CV_i^c|=~\sum_{l\in{[L]}}\min(\alpha,|\CU_i^l|)$, and the total number of possible ways to build $\CV_i^c$ is
\begin{equation}
    \prod_{l\in{[L]}} \binom{|\CU_i^l|}{\min(\alpha,|\CU_i^l|)}.
\end{equation}

Let us use $\CL(\CV_i^c)$ to denote the set of cache profiles assigned to all users in $\CV_i^c$. In order to build the messages to be transmitted by $h_i$, we first create the extended set of $\CV_i^c$, denoted by $\hat{\CV}_i^c$, by adding $l^c(i) = L - |\CL(\CV_i^c)|$ phantom users $u_{l}^*$ to $\CV_i^c$, such that $\SfL(u_l^*) = l$, for every $l \in [L] \backslash \CL(\CV_i^c)$. Then, 
%To find the messages to be transmitted by helper $h_i$ for the users in $\CV_i^c$, we create its \emph{extended} set $\hat{\CV}_i^c$ by adding $L - |\CL(\CV_i^c)|$ \emph{phantom} users $u_{l}^*$, $\SfL(u_l^*) = l$, for every $l \in [L] \backslash \CL(\CV_i^c)$. Next, 
for every subset $\CS$ of $[L]$ where $|\CS|=t+1$,\footnote{If $\alpha$ is very large, we can improve the performance further by choosing more than $t+1$ cache profiles in each transmission~\cite{abolpour2023cache}. This will be studied in the extended version of the paper.} we build a subset $\hat{\CS}_i^c$ of $\hat{\CV}_i^c$ that includes every user $u_k \in \hat{\CV}_i^c$ for which $\SfL(u_k) \in \CS$. Finally, for every resulting $\hat{\CS}_i^c$, we create the preliminary message\footnote{Note that we have used the term `message' instead of `codeword' here to distinguish the underlying signal-level interference cancellation mechanism~\cite{salehi2022enhancing}.} $\hat{X}(\hat{\CS}_i^c)$ as
\begin{equation}
\label{eq:codeowrd_ccc}
    \hat{X}(\hat{\CS}_i^c) = \sum_{u_{k} \in {\hat{\CS}_i^c}} W_{d_{k}, \CS \backslash \SfL(u_k) }\Bw_{\CM(\hat{\CS_i^c}), u_k} ,
\end{equation}
where the beamforming vector $\Bw_{\CM(\hat{\CS_i^c}), u_k}$ suppresses data at every user in
\begin{equation}
    \CM(\hat{\CS_i^c}), u_k = \hat{\CS}_i^{c,\SfL(u_k)} \backslash \{u_k\}
\end{equation}
where $\hat{\CS}_i^{c,\SfL(u_k)}$ denotes the set of all users in $\hat{\CS}_i^{c}$ that are assigned to the same cache profile as $u_k$. In other words, $\hat{\CS}_i^{c,\SfL(u_k)}$ includes every user $u_{\hat{k}} \in \hat{\CS}_i^{c}$ for which $\SfL(u_{\hat{k}}) = \SfL(u_k)$. The real transmitted message $X(\CS_i^c)$ is built from $\hat{X}(\hat{\CS}_i^c)$ by removing the effect of phantom users. 
It can be easily verified that every user $u_k$ in the feasible set $\CV_i^c$ can successfully decode their requested chunk after $ \Sfn(\CV_i^c) = {\binom{L}{t+1} - \binom{l(i)}{t+1}}$ transmissions. The rate vectors are then calculated similarly as in Section~\ref{section:siso}. 
%Note that the number of transmissions depends on $|\CL(\CV_i^c)|$ but not on $|\CV_i^c|$ as was the case for SISO setups.
%where $|\CL(\CV_i)|=|\CV_i|$ such that $l(i)=L-|\CV_i|$.

Note that with the message creation process in~\eqref{eq:codeowrd_ccc}, the interference from the data sent to a user $u_k$ is removed by cache contents at users with a cache profile different than $u_k$, and suppressed by beamforming at users with the same cache profile as $u_k$. This is a similar interference mitigation mechanism as in the shared-cache model~\cite{parrinello2019fundamental,parrinello2020extending}.

\begin{exmp}
Consider the same network and activation pattern as in Example~\ref{exmp:siso}. Let $t=1$ and $\alpha=2$. We have $\CU_2=\{u_3,u_4,u_5\}$ and $\CV_2^c=\CU_2$, and $\hat{\CV}_2^c = \{u_3,u_4,u_5,u_3^*\}$. There exist three subsets $\CS$ of $[L]$ with size $t+1$, 
%and three subsets $\hat{\CT}_2^c$ of $\hat{V}_2^c$ where  $\hat{\CT}_2^c=\{\hat{\CV}_2^{c,l}|$ s.t. $l\in{\CT}\}$ with $\CL({\hat{\CT}_2^c})=3$, 
resulting in preliminary messages
    %\begin{equation*}
        \begin{align*}
            \hat{X}(\{u_3,u_5,u_4\}) &= W_{d_3,\{1\}}\Bw_{\{5\}} +W_{d_5,\{1\}}\Bw_{\{3\}} + W_{d_4,\{2\}}, \\
            \hat{X}(\{u_3,u_5,u_3^*\}) &= W_{d_3,\{3\}}\Bw_{\{5\}} + W_{d_5,\{3\}}\Bw_{\{3\}} + W_{d_{3^*},\{2\}}, \\
            \hat{X}(\{u_4,u_3^*\}) &= W_{d_4,\{3\}} + W_{d_{3^*},\{1\}},
        \end{align*}
    %\end{equation*}
where $\Bw_{\{k\}}$ is the beamforming vector suppressing data at user $u_k$. Denoting the files requested by users $u_1$-$u_5$ with $A$-$E$, after removing the effect of phantom users, the real transmitted messages will be
$X(\{u_3,u_5,u_4\}) = C_1 \Bw_{\{5\}} + E_1 \Bw_{\{3\}} + D_2$, $X(\{u_3,u_5\}) = C_3 \Bw_{\{5\}} + E_3 \Bw_{\{3\}}$, and  $X(\{u_4\}) = D_3$.
%where $\Bz_k$ is zf-vector acting on user $u_k$ and $\Bz_0$ means there is no zero-forcing. 
%After removing the effect of phantom users, final vectors are built as 
%$X(\{u_3,u_5,u_4\}) = W_{d_3,\{1\}}\Bw_{\{5\}} +W_{d_5,\{1\}}\Bw_{\{3\}} + W_{d_4,\{2\}}$, $X(\{u_3,u_5\}) = W_{d_3,\{3\}}\Bw_{\{5\}} + W_{d_5,\{3\}}\Bw_{\{3\}}$, and  $X(\{u_4\})=W_{d_4,\{3\}}$. 
The resulting rate vector is then given as $\Br(j,s) = [0,0,\frac{1}{3},\frac{1}{3},\frac{1}{3},0]$. %and it is easy to verify that for $\{u_3,u_4\}$, $\{u_3,u_5\}$ and $\{u_3\}$ the rate vectors are $[0,0,\frac{1}{3},\frac{1}{3},0,0]$, $[0,0,\frac{1}{2},0,\frac{1}{2},0]$ and $[0,0,\frac{1}{2},0,0,0]$ respectively.
\end{exmp}

%For simplicity and for the sake of this paper as in Section ..., we restrict ourselves to maximal rate vectors that maximize the coded caching gain for each helper, i.e. the rate vectors which have maximum $\CL(\CV_i)$, so we have $\CL(\CV_i)=L-l(i)$ such that the rates of users in $\CV_i$ become $r_{\Sfk(i)} = {1}/({\binom{L}{t+1}-\binom{\Sfl(i)}{t+1}})$ as in Section .... It is easy to see that number of policies per activation pattern $\Bp_j$ gets upper bounded by
%\begin{equation}
%\label{eq:activation_pattern_count_maximal}
%    \Sfc(\Bp_j) \leq \prod_{\substack{i \in [H]\\ \Bp_j[i] = 1}} \prod_{l \in [L]} \binom{\max(|\CU_i^l|,\alpha)}{\alpha} .
%\end{equation}

%\subsection{Scheduling with Interference Cancellation}

%% file: SimResults.tex
We use numerical results to compare the performance of SISO and MISO transmission schemes. We assume $H$ helpers are located at the center of hexagons on a limited hexagonal grid. Every hexagon has a radius of~$1$, and we have $r_{\mathrm{trans}} = 1$ and $r_{\mathrm{inter}}  =1.2$. The users are placed according to a homogeneous Poisson Point Process in the transmission area of helpers, and there is at least one helper within the transmission radius of every user. The average number of users per helper, shown by $U$, is determined by the density of the Poisson Point process. Every user is randomly assigned with a cache profile $l \in [L]$. For all MISO setups, we have set $\alpha=2$.

Simulation results are provided in Figures~\ref{fig:cdf_l=all}-\ref{fig:cdf_l=6}. In the figures, $L=1$ corresponds to the case where no CC technique is applied, i.e., cache contents are used only locally.\footnote{In fact, with $L=1$, interference reduction and cache congestion control are similar to inter- and intra-cell (helper) interference removal, respectively.}
Moreover, the `optimum selection' strategy for MISO systems selects the best performance among interference reduction and cache congestion control methods, i.e., all the rate vectors of both methods are used while solving~\eqref{eq:optimization_problem_main}.\footnote{in practice, the optimal strategy can be the `default' approach if the (moderate) increase in the computation complexity of solving the optimization problem~\eqref{eq:optimization_problem_main} is acceptable.}
%, which makes sense as the scheduler can choose either of the methods for each scheduling decision). %(upper bound). %The users are placed in the vicinity of helpers according to a homogeneous Poisson Point Process, and $U$ denotes the average number of users per helper. For all MISO setups, we have set $\alpha = 2$.

As can be seen in Figure~\ref{fig:cdf_l=all}, MISO setups provide much better performance than SISO, which is expected as they benefit from the extra resources provided by multi-antenna transmitters. Moreover, from Figures~\ref{fig:cdf_l=1}-\ref{fig:cdf_l=6}, we see that the cache congestion control method generally outperforms interference reduction. This is because the underlying shared cache model of the cache congestion control method has a larger degree of freedom (DoF), enabling more users to be served simultaneously~\cite{parrinellomultiantenna,parrinello2020extending}. 
%\color{blue} To see the DoF difference, notice that in the interference nulling method, every active helper in an activation pattern can suppress its transmission at a maximum of $\alpha-1$ users, i.e., free $\alpha-1$ users to be served by other helpers, whereas in the cache congestion control method, every active helper can potentially serve extra $L(\alpha-1)$ users. \color{black} 
However, as the $\frac{U}{L}$ ratio decreases, there is less chance of finding users with the same cache contents, and the performance gap between the two methods narrows. 
%In fact, as can be seen from Figure~\ref{fig:cdf_l=6}, when $L=U$, in some realizations, the interference nulling method provides superior performance. 

One particular behavior in Figure~\ref{fig:cdf_l=all} is for the optimum selection strategy, where the $L=3$ case outperforms $L=6$. To understand this behavior, we note that the optimum selection strategy performs close to the cache congestion control method (see Figure~~\ref{fig:cdf_l=3}), which actually performs better for the network setup considered in Figure~\ref{fig:cdf_l=all} if $L=3$. This is because if $U=6$ and $L=3$, on average, every cache profile will be assigned to two users, which is the ideal case for the cache congestion control method when $\alpha = 2$. Using a similar intuition, in the general case, we expect that the cache congestion control mechanism to provide its best performance when $L\approx \frac{U}{\alpha}$.

\color{black}

\input{Figs-MISO/Plot-Comp}
\input{Figs-MISO/Plot-L1}
\input{Figs-MISO/Plot-L3}
\input{Figs-MISO/Plot-L6}

%\todo[inline]{Revised up to here.}

%We use numerical results to compare the performance of various solutions. We assume $H$ helpers are located at the center of hexagons on a limited hexagonal grid. Every hexagon has a radius of~$1$, and we have $r_{\mathrm{trans}} = 1$ and $r_{\mathrm{inter}}  =1.2$. The users are placed according to a homogeneous Poisson Point Process in the transmission area of helpers. The average number of users per helper, shown by $U$, is determined by the density of the Poisson process. Every user is randomly assigned with a cache profile $l \in [L]$.

%% file: Figs-MISO/Plot-Comp.tex
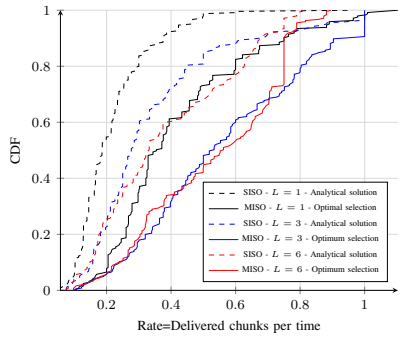
\begin{figure}[t]
    \centering
    \resizebox{0.60\columnwidth}{!}{%
    
    \begin{tikzpicture}

    \begin{axis}
    [
    % put axis lines at left and bottom
    axis lines = left,
    % control axis labels
    xlabel = \smaller {Rate=Delivered chunks per time},
    ylabel = \smaller {CDF},
    ylabel near ticks,
    % control legend position
    legend pos = south east,
    % control size of tick marks (10,20,30,etc)
    ticklabel style={font=\smaller},
    % control major grids
    grid=both,
    major grid style={line width=.2pt,draw=gray!30},
    % control minor grids
    %grid style={line width=.1pt, draw=gray!10},
    %minor tick num=5,
    ]
    
    % \addplot[black]
    % table[y=Opt-L1-Y,x=Opt-L1-X]{Figs/CDF_data_L.tex};
    % \addlegendentry{\smaller $L=1$, Analytical}
    % \addplot[black,dashed]
    % table[y=Super-L1-Y,x=Super-L1-X]{Figs/CDF_data_L.tex};
    % \addlegendentry{\smaller $L=1$, Super-user}
    % \addplot[black!50]
    % table[y=Heur-L1-Y,x=Heur-L1-X]{Figs/CDF_data_L.tex};
    % \addlegendentry{\smaller $L=1$, RGA}
    
    % \addplot[green]
    % table[y=Opt-L5-Y,x=Opt-L5-X]{Figs/CDF_data_L.tex};
    % \addlegendentry{\smaller $L=5$, Analytical}
    % \addplot[green,dashed]
    % table[y=Super-L5-Y,x=Super-L5-X]{Figs/CDF_data_L.tex};
    % \addlegendentry{\smaller $L=5$, Super-user}
    % \addplot[green!50]
    % table[y=Heur-L5-Y,x=Heur-L5-X]{Figs/CDF_data_L.tex};
    % \addlegendentry{\smaller $L=5$, RGA}
    
    \addplot[black,dashed]
    table[y=CDF-L1-A1,x=Rate-L1-A1]{Figs-MISO/MISO_Data_April27.tex};
    \addlegendentry{\tiny SISO - $L=1$ - Analytical solution}
    \addplot[black]
    table[y=CDF-L1-A2-OPT,x=Rate-L1-A2-OPT]{Figs-MISO/MISO_Data_April27.tex};
    \addlegendentry{\tiny MISO - $L=1$ - Optimal selection}
    \addplot[blue,dashed]
    table[y=CDF-L3-A1,x=Rate-L3-A1]{Figs-MISO/MISO_Data_April27.tex};
    \addlegendentry{\tiny SISO - $L=3$ - Analytical solution}
    \addplot[blue]
    table[y=CDF-L3-A2-OPT,x=Rate-L3-A2-OPT]{Figs-MISO/MISO_Data_April27.tex};
    \addlegendentry{\tiny MISO - $L=3$ - Optimum selection}
    \addplot[red,dashed]
    table[y=CDF-L6-A1,x=Rate-L6-A1]{Figs-MISO/MISO_Data_April27.tex};
    \addlegendentry{\tiny SISO - $L=6$ - Analytical solution}
    \addplot[red]
    table[y=CDF-L6-A2-OPT,x=Rate-L6-A2-OPT]{Figs-MISO/MISO_Data_April27.tex};
    \addlegendentry{\tiny MISO - $L=6$ - Optimum selection}

    \end{axis}

    \end{tikzpicture}
    }
    \vspace{-8pt}
    \caption{CDF of the user rates, $H=4$, $U=6$, $\gamma = 1/3$.
    %, time is calculated assuming a full video chunk can be transmitted in one second.
    }
    \label{fig:cdf_l=all}
    \vspace{-10pt}
\end{figure}

%% file: Figs-MISO/Plot-L1.tex
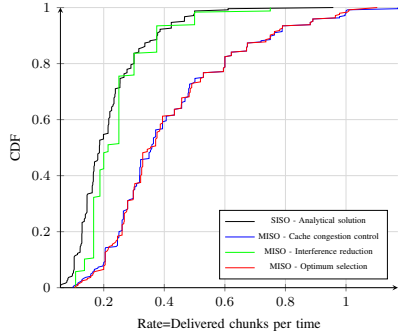
\begin{figure}[t]
    \centering
    \resizebox{0.60\columnwidth}{!}{%
    
    \begin{tikzpicture}

    \begin{axis}
    [
    % put axis lines at left and bottom
    axis lines = left,
    % control axis labels
    xlabel = \smaller {Rate=Delivered chunks per time},
    ylabel = \smaller {CDF},
    ylabel near ticks,
    % control legend position
    legend pos = south east,
    % control size of tick marks (10,20,30,etc)
    ticklabel style={font=\smaller},
    % control major grids
    grid=both,
    major grid style={line width=.2pt,draw=gray!30},
    % control minor grids
    %grid style={line width=.1pt, draw=gray!10},
    %minor tick num=5,
    ]
    
    % \addplot[black]
    % table[y=Opt-L1-Y,x=Opt-L1-X]{Figs/CDF_data_L.tex};
    % \addlegendentry{\smaller $L=1$, Analytical}
    % \addplot[black,dashed]
    % table[y=Super-L1-Y,x=Super-L1-X]{Figs/CDF_data_L.tex};
    % \addlegendentry{\smaller $L=1$, Super-user}
    % \addplot[black!50]
    % table[y=Heur-L1-Y,x=Heur-L1-X]{Figs/CDF_data_L.tex};
    % \addlegendentry{\smaller $L=1$, RGA}
    
    % \addplot[green]
    % table[y=Opt-L5-Y,x=Opt-L5-X]{Figs/CDF_data_L.tex};
    % \addlegendentry{\smaller $L=5$, Analytical}
    % \addplot[green,dashed]
    % table[y=Super-L5-Y,x=Super-L5-X]{Figs/CDF_data_L.tex};
    % \addlegendentry{\smaller $L=5$, Super-user}
    % \addplot[green!50]
    % table[y=Heur-L5-Y,x=Heur-L5-X]{Figs/CDF_data_L.tex};
    % \addlegendentry{\smaller $L=5$, RGA}
    
    \addplot[black]
    table[y=CDF-L1-A1,x=Rate-L1-A1]{Figs-MISO/MISO_Data_April27.tex};
    \addlegendentry{\tiny SISO - Analytical solution}
    \addplot[blue]
    table[y=CDF-L1-A2-CC,x=Rate-L1-A2-CC]{Figs-MISO/MISO_Data_April27.tex};
    \addlegendentry{\tiny MISO - Cache congestion control}
    \addplot[green]
    table[y=CDF-L1-A2-IC,x=Rate-L1-A2-IC]{Figs-MISO/MISO_Data_April27.tex};
    \addlegendentry{\tiny MISO - Interference reduction}
    \addplot[red]
    table[y=CDF-L1-A2-OPT,x=Rate-L1-A2-OPT]{Figs-MISO/MISO_Data_April27.tex};
    \addlegendentry{\tiny MISO - Optimum selection}
    %\addplot[gray]
    %table[y=Opt-L20-Y,x=Opt-L20-X]{Figs/CDF_data_L.tex};
    %\addlegendentry{\smaller $L=20$}

    \end{axis}

    \end{tikzpicture}
    }
    \vspace{-8pt}
    \caption{CDF of the user rates, $H=4$, $U=6$, $L=1$, $\gamma = 1/3$.
    %, time is calculated assuming a full video chunk can be transmitted in one second.
    }
    \label{fig:cdf_l=1}
    \vspace{-10pt}
\end{figure}

%% file: Figs-MISO/Plot-L3.tex
\begin{figure}[ht]
    \centering
    \resizebox{0.60\columnwidth}{!}{%
    
    \begin{tikzpicture}

    \begin{axis}
    [
    % put axis lines at left and bottom
    axis lines = left,
    % control axis labels
    xlabel = \smaller {Rate=Delivered chunks per time},
    ylabel = \smaller {CDF},
    ylabel near ticks,
    % control legend position
    legend pos = south east,
    % control size of tick marks (10,20,30,etc)
    ticklabel style={font=\smaller},
    % control major grids
    grid=both,
    major grid style={line width=.2pt,draw=gray!30},
    % control minor grids
    %grid style={line width=.1pt, draw=gray!10},
    %minor tick num=5,
    ]
    
    % \addplot[black]
    % table[y=Opt-L3-Y,x=Opt-L1-X]{Figs/CDF_data_L.tex};
    % \addlegendentry{\smaller $L=1$, Analytical}
    % \addplot[black,dashed]
    % table[y=Super-L1-Y,x=Super-L1-X]{Figs/CDF_data_L.tex};
    % \addlegendentry{\smaller $L=1$, Super-user}
    % \addplot[black!50]
    % table[y=Heur-L1-Y,x=Heur-L1-X]{Figs/CDF_data_L.tex};
    % \addlegendentry{\smaller $L=1$, RGA}
    
    % \addplot[green]
    % table[y=Opt-L5-Y,x=Opt-L5-X]{Figs/CDF_data_L.tex};
    % \addlegendentry{\smaller $L=5$, Analytical}
    % \addplot[green,dashed]
    % table[y=Super-L5-Y,x=Super-L5-X]{Figs/CDF_data_L.tex};
    % \addlegendentry{\smaller $L=5$, Super-user}
    % \addplot[green!50]
    % table[y=Heur-L5-Y,x=Heur-L5-X]{Figs/CDF_data_L.tex};
    % \addlegendentry{\smaller $L=5$, RGA}
    
    \addplot[black]
    table[y=CDF-L3-A1,x=Rate-L3-A1]{Figs-MISO/MISO_Data_April27.tex};
    \addlegendentry{\tiny SISO - Analytical solution}
    \addplot[blue]
    table[y=CDF-L3-A2-CC,x=Rate-L3-A2-CC]{Figs-MISO/MISO_Data_April27.tex};
    \addlegendentry{\tiny MISO - Cache congestion control}
    \addplot[green]
    table[y=CDF-L3-A2-IC,x=Rate-L3-A2-IC]{Figs-MISO/MISO_Data_April27.tex};
    \addlegendentry{\tiny MISO - Interference reduction}
    \addplot[red]
    table[y=CDF-L3-A2-OPT,x=Rate-L3-A2-OPT]{Figs-MISO/MISO_Data_April27.tex};
    \addlegendentry{\tiny MISO - Optimum selection}
    %\addplot[gray]
    %table[y=Opt-L20-Y,x=Opt-L20-X]{Figs/CDF_data_L.tex};
    %\addlegendentry{\smaller $L=20$}

    \end{axis}

    \end{tikzpicture}
    }
    \vspace{-8pt}
    \caption{CDF of the user rates, $H=4$, $U=6$, $L=3$, $\gamma = 1/3$.
    %, time is calculated assuming a full video chunk can be transmitted in one second.
    }
    \label{fig:cdf_l=3}
    \vspace{-10pt}
\end{figure}
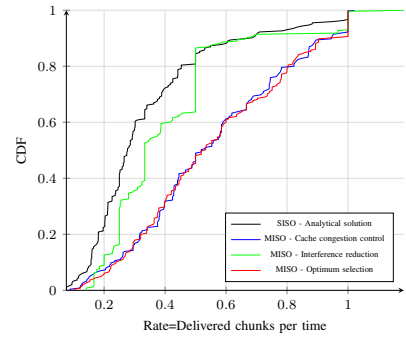

%% file: Figs-MISO/Plot-L6.tex
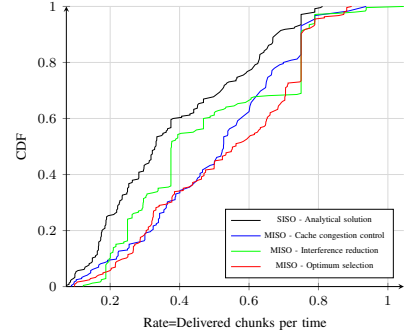
\begin{figure}[ht]
    \centering
    \resizebox{0.60\columnwidth}{!}{%
    
    \begin{tikzpicture}

    \begin{axis}
    [
    % put axis lines at left and bottom
    axis lines = left,
    % control axis labels
    xlabel = \smaller {Rate=Delivered chunks per time},
    ylabel = \smaller {CDF},
    ylabel near ticks,
    % control legend position
    legend pos = south east,
    % control size of tick marks (10,20,30,etc)
    ticklabel style={font=\smaller},
    % control major grids
    grid=both,
    major grid style={line width=.2pt,draw=gray!30},
    % control minor grids
    %grid style={line width=.1pt, draw=gray!10},
    %minor tick num=5,
    ]
    
    % \addplot[black]
    % table[y=Opt-L1-Y,x=Opt-L1-X]{Figs/CDF_data_L.tex};
    % \addlegendentry{\smaller $L=1$, Analytical}
    % \addplot[black,dashed]
    % table[y=Super-L1-Y,x=Super-L1-X]{Figs/CDF_data_L.tex};
    % \addlegendentry{\smaller $L=1$, Super-user}
    % \addplot[black!50]
    % table[y=Heur-L1-Y,x=Heur-L1-X]{Figs/CDF_data_L.tex};
    % \addlegendentry{\smaller $L=1$, RGA}
    
    % \addplot[green]
    % table[y=Opt-L5-Y,x=Opt-L5-X]{Figs/CDF_data_L.tex};
    % \addlegendentry{\smaller $L=5$, Analytical}
    % \addplot[green,dashed]
    % table[y=Super-L5-Y,x=Super-L5-X]{Figs/CDF_data_L.tex};
    % \addlegendentry{\smaller $L=5$, Super-user}
    % \addplot[green!50]
    % table[y=Heur-L5-Y,x=Heur-L5-X]{Figs/CDF_data_L.tex};
    % \addlegendentry{\smaller $L=5$, RGA}
    
    \addplot[black]
    table[y=CDF-L6-A1,x=Rate-L6-A1]{Figs-MISO/MISO_Data_April27.tex};
    \addlegendentry{\tiny SISO - Analytical solution}
    \addplot[blue]
    table[y=CDF-L6-A2-CC,x=Rate-L6-A2-CC]{Figs-MISO/MISO_Data_April27.tex};
    \addlegendentry{\tiny MISO - Cache congestion control}
    \addplot[green]
    table[y=CDF-L6-A2-IC,x=Rate-L6-A2-IC]{Figs-MISO/MISO_Data_April27.tex};
    \addlegendentry{\tiny MISO - Interference reduction}
    \addplot[red]
    table[y=CDF-L6-A2-OPT,x=Rate-L6-A2-OPT]{Figs-MISO/MISO_Data_April27.tex};
    \addlegendentry{\tiny MISO - Optimum selection}
    %\addplot[gray]
    %table[y=Opt-L20-Y,x=Opt-L20-X]{Figs/CDF_data_L.tex};
    %\addlegendentry{\smaller $L=20$}

    \end{axis}

    \end{tikzpicture}
    }
    \vspace{-8pt}
    \caption{CDF of the user rates, $H=4$, $U=6$, $L=6$, $\gamma = 1/3$.
    %, time is calculated assuming a full video chunk can be transmitted in one second.
    }
    \label{fig:cdf_l=6}
    \vspace{-10pt}
\end{figure}